\documentclass[a4paper,fleqn,usenatbib]{mnras}

\usepackage[T1]{fontenc}
\usepackage{ae,aecompl}

\usepackage{graphicx}	% Including figure files
\usepackage{amsmath}	% Advanced maths commands
\usepackage{amssymb}	% Extra maths symbols

\newcommand{\beq}{\begin{equation}}
\newcommand{\eeq}{\end{equation}}
\newcommand{\ba}{\begin{eqnarray}}
\newcommand{\ea}{\end{eqnarray}}
\newcommand{\dd}{\mathrm{d}}

\newcommand{\eps}{{\epsilon}}

\newcommand{\Sg}{{\Sigma}}
\newcommand{\lam}{{\lambda}}
\newcommand{\w}{{\omega}}

\newcommand{\Gm}{{\Gamma}}

\newcommand{\dt}{{\dd t}}
\newcommand{\dr}{{\dd r}}
\newcommand{\dphi}{{\dd \phi}}

\newcommand{\dlam}{{\dd \lam}}

\newcommand{\df}{{\dd f}}
\newcommand{\drho}{{\dd \rho}}
\newcommand{\dz}{{\dd z}}

\newcommand{\IB}{\mathfrak{B}}
\newcommand{\gt}{\tilde{g}(r)}
\newcommand{\gtw}{\tilde{g}(\w)}

\newcommand{\IBw}{\IB_{n q \varrho} (\w)}
\newcommand{\IBf}{\IB_{n q \varrho} (f)}

\newcommand{\Mc}{\mathcal{M}}
\newcommand{\rspc}{\Upsilon}
\newcommand{\rmin}{\rspc_{\rm max_{H}}}
\newcommand{\rLR}{\rspc_{\rm max_{LR}}}

\title[Testing pc-GR with gravitational waves]{Testing pseudo-complex general relativity with gravitational waves}

\author[Nielsen et al.]{
Alex B. Nielsen,$^{1,2}$\thanks{E-mail: alex.nielsen@aei.mpg.de}
Ofek Birnholz,$^{1,2}$
\\
$^{1}$Albert-Einstein-Institut, Max-Planck-Institut f\"ur Gravi\-ta\-tions\-physik, D-30167 Hannover, Germany\\
$^{2}$Leibniz Universit\"at Hannover, D-30167 Hannover, Germany
}

\date{Accepted XXX. Received YYY; in original form ZZZ}

\pubyear{2017}

\begin{document}
\label{firstpage}
\pagerange{\pageref{firstpage}--\pageref{lastpage}}
\maketitle

\begin{abstract}
We show how the model of pseudo-complex general relativity can be tested
using gravitational wave signals from coalescing compact objects.
The model, which agrees with Einstein gravity in the weak-field limit,
	diverges dramatically in the near-horizon regime,
	with certain parameter ranges excluding the existence of black holes.
We show that simple limits can be placed on the model in both the inspiral and ringdown phase of coalescing compact objects.
We discuss further how these bounds relate to current observational limits.
In particular, for minimal scenarios previously considered in the literature,
	gravitational wave observations are able to constrain pseudo-complex general relativity parameters
	to values that require the existence of black hole horizons.
\end{abstract}

\begin{keywords}
gravitational waves -- black hole physics -- methods: observational -- relativistic processes -- binaries: general
\end{keywords}

%%%%%%%%%%%%%%%%%%%%%%%%%%%%%%%%%%%%%%%%%%%%%%%%%%

%%%%%%%%%%%%%%%%% BODY OF PAPER %%%%%%%%%%%%%%%%%%

\section{Introduction}

Gravitational wave detections of coalescing binary black holes
	(\cite{TheLIGOScientific:2016pea, Abbott:2016blz, Abbott:2016nmj, Abbott:2017vtc})
	with the Advanced LIGO detectors
	(\cite{Abramovici:1992ah, Harry:2010zz, Aasi:2013wya, TheLIGOScientific:2014jea, TheLIGOScientific:2016agk})
	have opened up a new tool to test modified theories of gravity.
A number of such tests have already been performed looking for generic deviations from Einstein's general relativity (GR),
	and so far the data has been found consistent with GR
	(\cite{TheLIGOScientific:2016pea, Abbott:2016blz, TheLIGOScientific:2016src, Abbott:2017vtc, Abbott:2016bqf}).
Merging black hole observations are particularly suited to testing theories that deviate from Einstein's GR in the near black hole horizon regime.
In a certain sense, colliding black holes are the ideal testing ground for such models,
	but because of a lack of definite predictions, little is known about how these tests impact specific modified theories.

One such theory that proposes to modify the near-horizon physics is pseudo-complex general relativity (pc-GR) (\cite{Hess:2008wd}).
This theory proposes a pseudo-complex generalisation of GR and leads to a number of novel predictions.
Ray-tracing of light rays in this geometry has been previously calculated in \cite{Schonenbach:2013nya}
	with an eye to comparing predictions to observations from the Event Horizon Telescope.
Simulations of accretion disks have been studied to compare to X-ray observations of accreting systems (\cite{Hess:2015hpc}),
associated with an innermost stable circular orbit (ISCO).
Further studies have looked at the effect on gravitational redshift and frame-dragging (\cite{Schonenbach:2012mw}).
Most of these studies relate to comparisons with future precision observational data.

One of the features believed to be associated with pc-GR is a strong modification of near-horizon physics.
In fact it is has been claimed that pc-GR predicts there are no black holes
	because of a modification of the near-horizon gravitational field (\cite{Hess:2010pba}).
This makes gravitational wave observations of merging black holes ideal observations to test such a theory.
Here we will directly compare the theory with the gravitational wave observation of GW150914
	(\cite{Abbott:2016blz, Abbott:2016bqf})
	and with bounds on modifications from GR established by Advanced LIGO's first detections
	(\cite{TheLIGOScientific:2016pea, Abbott:2016blz, TheLIGOScientific:2016src, Abbott:2016nmj, Abbott:2017vtc}).
This has previously been investigated in \cite{Hess:2016gmh}, where it was claimed that
	pc-GR implies that the coalescence which produced GW150914
	may have had a chirp mass much larger than claimed by the LIGO team,
	and occurring at a much greater luminosity distance.
Here we will re-examine this interpretation and show how existing gravitational wave observations
	are able to constrain the free parameters of pc-GR
	and even rule out certain parameter ranges that allow horizonless objects.

\section{Model and spacetime metric}

In any metric theory of gravity, including pc-GR, the spacetime of an isolated, spinning, stationary object is likely to be axisymmetric.
By the requirements of theorem 7.1.1 of (\cite{wald2010general}) such a stationary, axisymmetric metric can be put in the form
\beq \label{generalKerr}
\dd s^2 = g_{tt}\dt^2 + 2g_{\phi t} \dt \dphi + g_{\phi\phi}\dphi^2 + g_{\rho\rho}\drho^{2} + g_{zz}\dz^{2} ~,
\eeq
where only four of the metric functions are independent since $\rho^{2} = g^2_{t\phi} - g_{tt}g_{\phi\phi}$,
	and all metric functions are only functions of the coordinates $\rho$ and $z$.
The coordinates $t$ and $\phi$ are adapted to the stationary and axisymmetric symmetries and the coordinate $z$ can be replaced with a zenith-angle coordinate $\theta$.
If we furthermore assume that the spacetime has a reflection symmetry about an equatorial plane, $\theta=\frac{\pi}{2}$,
	then all metric functions are guaranteed to satisfy $\partial_{\theta}g_{ab}=0$ on this equatorial plane.
This general form of the metric will hold in any metric theory of gravity with these symmetries
	and is a purely geometrical result, before any theory-specific equations of motion have been solved.
In particular,
	it contains the Kerr-Newman class of electrovacuum solutions in GR as well as a host of other known solutions.

To obtain the functional form of the metric functions in a specific theory we should solve the equations of motion.
This has already been done within pc-GR (\cite{Caspar:2012ux}) and we adopt without modification the solution here:
\ba
g_{tt} &=& - \left( 1 - \frac{\psi}{\Sg} \right)		~~~,	~~~%\nonumber \\
g_{rr} = \frac{\Sg}{\Delta}		~~~,	~~~ % \nonumber \\
g_{\theta\theta} = \Sg		~,	\nonumber \\
g_{\phi\phi} &=& \left( \left( r^2 + a^2 \right) + \frac{a^2\psi}{\Sg} \sin^2\theta \right) \sin^2\theta			~,	\nonumber \\
g_{t\phi} &=& g_{\phi t} = - a \frac{\psi}{\Sg} \sin^2\theta		~,
\label{metric}
\ea
with $\Sg = r^2 + a^2 \cos^2 \theta$ and $\Delta = r^2 + a^2 - \psi(r)$.
If the function $\psi(r)$ is chosen to satisfy $\psi =2Mr$
	then this solution is just the Kerr solution of vacuum GR with mass $M$ and specific angular momentum $a = \chi M$.
However, pc-GR allows for $\psi(r)$ to be a more general function,
	with the form adopted in \cite{Caspar:2012ux} $\psi  = 2 m(r) r$ where
\ba
m(r) = M - \frac{B}{2r^{n}} = M \, g(r)	~,	~~
g(r) = \left[ 1 - b \left(\frac{M}{r}\right)^{n} \, \right] .
\label{modified mass}
\ea
Here $b$ is a new dimensionless parameter for the pc-GR modification. Its value in GR is zero.
The ultimate provenance of this free parameter in pc-GR is a term in the action variation for the theory,
that following a proposal of \cite{Schuller:2002ma} is taken to lie exclusively in one half of the kinematical pseudo-complex algebra.
The freedom to choose $\psi(r)$ is a restricted form of the four metric-function freedom in the more general metric (\ref{generalKerr}).
In fact, in the non-rotating case with $g_{t\phi}=0$,
these solutions are examples of a restricted class of dirty black holes studied previously in Einstein's GR (\cite{Visser:1992qh});
	of particular relevance to the current work is their quasi-normal mode behaviour (\cite{Medved:2003rga}).
We will not discuss further here the theoretical motivation for including such a general function,
but merely adopt this as a model to be constrained by data.

The parameter $b$ can be chosen large enough such that the solutions do not admit Killing horizons,
although values less than this critical value are not in principle ruled out by the theory.
The Killing horizons of the Killing vector field
$k^{a} = \delta^{a}_{t} - \Omega\delta^{a}_{\phi}$
occur at the solutions of $r^{2}+a^{2}-2m(r)r =0$ where $\Omega = g_{t\phi}/g_{\phi\phi}$. 
No horizons exist when this equation does not have real positive solutions, which occurs when $b$ is greater than
\beq b_{\mathrm{maxH}} = \rmin^{n}\left( 1 - \frac{\chi^{2}}{2\rmin} - \frac{\rmin}{2}\right)  ~, \label{bmaxH} \eeq
where $\rmin = (n+\sqrt{n^2-(n^2-1)\chi^2})/(n+1)$.
Thus for sufficiently large $b$, black holes can be said not to exist (\cite{Hess:2010pba}).
This limiting value is largest when $\chi =0$, and in the $n=2$ case takes the value $16/27$.

As written, the spacetime metric is still in general singular,
and for $a=0$ contains a curvature singularity at $r=0$,
as evidenced by the value of the Kretschmann scalar
\ba
\frac{r^{6}}{4}R_{abcd}R^{abcd} = \hspace{5cm} & & \nonumber \\
	m'' r^{2} (4m-4m'r +r^{2}m'') + 8m'r(m'r-2m)+12m^{2} ~. & &
\ea
It is clearly the intention of the original authors of the model
that such singularities should be regularised by some effect (\cite{HessPrivate})
	and we take the spacetime of equation (\ref{metric}) as a working model only away from such singularities.

Equatorial circular orbits in the general spacetime (\ref{generalKerr}) have four-velocities, $u^{a}$, given by
\beq
u^{a} = \frac{\dt}{\dlam}\delta^{a}_{t} + \frac{\dphi}{\dlam}\delta^{a}_{\phi}	~,
\eeq
where $\lam$ parameterises the orbital path and can be chosen to be the proper time in the case of timelike orbits.
For these orbits to be geodesic, bound only by gravity,
we should in addition solve the geodesic equation $u^{a}\nabla_{a}u^{b} = 0$.
In the spacetime (\ref{generalKerr}), the only non-trivial equation of the four geodesic equations is the $r$-component.
This condition suffices to determine the functions $\frac{\dt}{\dlam}$ and $\frac{\dphi}{\dlam}$
	up to an overall normalisation as a function of the $r$ coordinate.
Since the orbital frequency observed asymptotically is given by $\w = \dphi / \dt$, the geodesic equation gives
\beq \label{omega_geodesic}
\w_{\pm} = 
\frac{-g_{t\phi}' \pm \sqrt{g_{t\phi}'^{2}-g_{\phi\phi}' g_{tt}'}}{g_{\phi\phi}'}	~,
\eeq
where $'$ denotes an $r$-derivative.
Therefore, using the metric components of (\ref{metric}) the geodesic equation can be written as
\beq \label{mgeodesic} (\w \, a - 1)^2(m - m' r) - \w^2 \, r^3 = 0 ~, \eeq
In the limit of $m'=0$ this gives the expected behaviour for the Kerr spacetime (\cite{Bardeen:1972fi}),
	and in the further Schwarzschild limit of $a=0$ it reduces to the familiar Kepler-like relation between $r$ and $\w$ (\cite{Abbott:2016bqf}).

\section{Post-merger: ringdown}

The single body metric (\ref{metric}) lends itself immediately to calculations of test mass orbital properties,
	needed for studies of ray tracing, accretion disks or ringdown frequencies.
Objects compact enough to support circular photon orbits are expected
	to have ringdown frequencies approximated by the gravitational wave frequency
	of a massless test particle at this ```light ring" (LR) (\cite{Cardoso:2016rao}).
For the orbits discussed above to be null, we require additionally that $u^{a}u_{a}=0$.
This is equivalent to requiring
\beq \label{omega_null}
\w_{\pm} =  \frac{-g_{t\phi} \pm \sqrt{g_{t\phi}^{2}-g_{\phi\phi} g_{tt}}}{g_{\phi\phi}}	~.
\eeq
Using the metric components of (\ref{metric}) the null condition can be rewritten as
\beq \label{mnull} 2m - r - 4\w ma + \w^2 (r^3 + ra^2 + 2ma^2) = 0 ~. \eeq
For a circular orbit to be both geodesic and null requires both eqns (\ref{mgeodesic}) and (\ref{mnull}) to be satisfied.
Eliminating $\w$ from these equations gives the location of the light ring.
This will be the root of a polynomial in $r$, in the present case given by
\beq \sqrt{\Delta}(r^3-a^2 F) + a(2r^2m + (r^2+a^2)F) - r\sqrt{rF}g_{\phi\phi} = 0 ~,\eeq
where $\Delta$ and $g_{\phi\phi}$ are functions given in eqn (\ref{metric}) and we have defined $F = m - m' r$ as in \cite{Hess:2016gmh}.
This equation will have a positive real root (and hence a light ring will exist) if its value at a local minimum is negative for positive $r$.
For the static, spherically symmetric case with $-g_{tt}=g_{rr}^{-1} = 1-2m(r)/r$ and $g_{t\phi} = 0$ the equation becomes
\beq r-3m(r) + m' r = 0 ~, \eeq
For the Schwarzschild solution with $m'=0$ this gives the familiar photon sphere at $r=3M$.
For the mass function (\ref{modified mass}), the existence of a light ring requires $b$ to be less than
\beq b_{\mathrm{maxLR}} = \frac{1}{n(3+n)}\left( \rLR\right)^{n+1} ~, \label{bmaxLR} \eeq
where $\rLR=3n/(n+1)$.
Thus for the non-rotating case, when $n>0$, if there is a horizon then there is also a photon sphere.

\begin{figure}
	\includegraphics[width=\columnwidth]{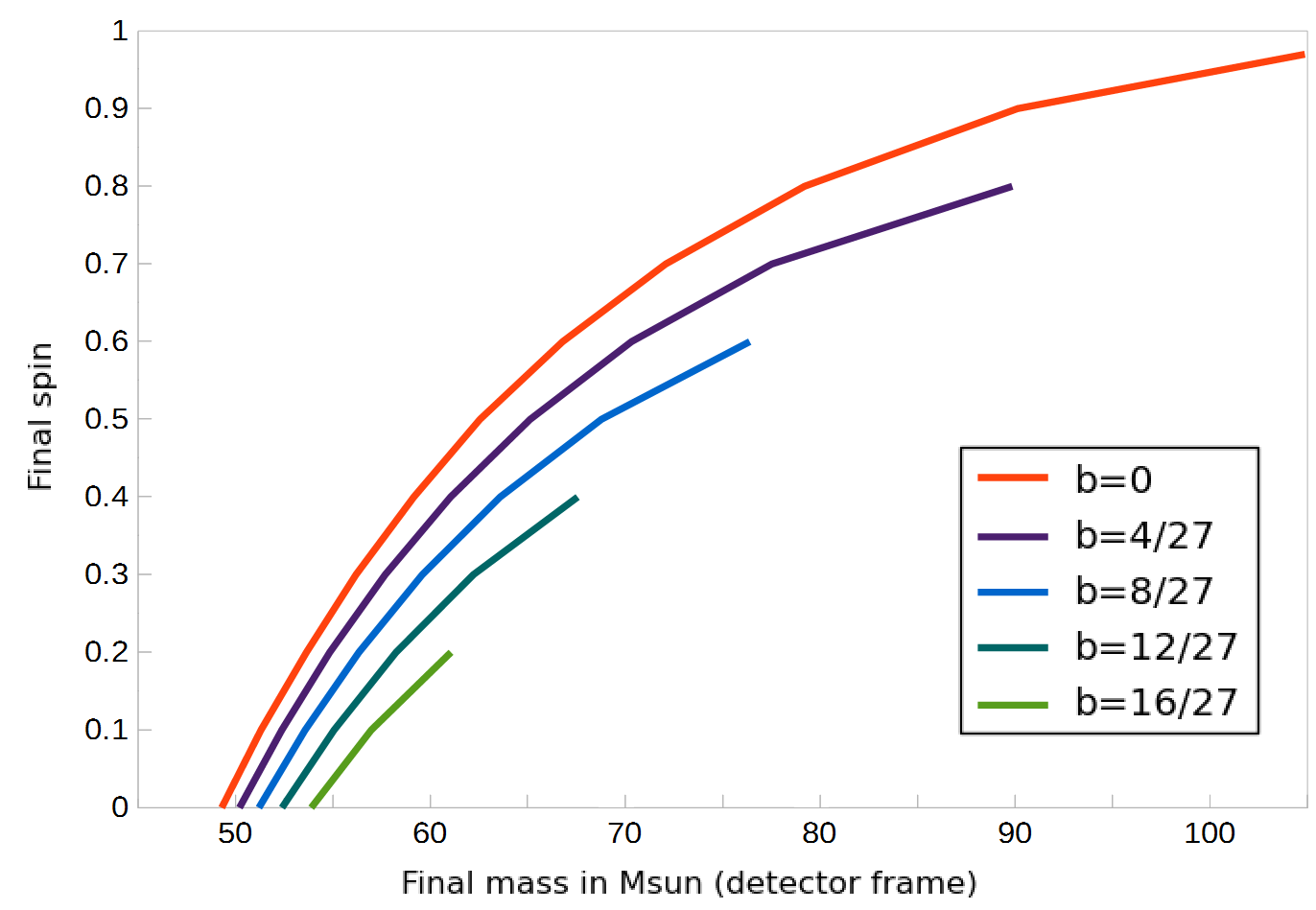}
    \caption{The final mass and spin values needed for the ringdown frequency, based on the light ring orbital frequency to take the value 250 Hz.
    Values of $b$ greater than zero require either a higher mass or a lower spin than the $b=0$ values predicted by GR.}
    \label{fig:ringdown}
\end{figure}

With the location of the light ring,
	the frequency of an orbiting null test mode can be calculated using either eqn (\ref{omega_geodesic}) or eqn (\ref{omega_null}).
Fig (\ref{fig:ringdown}) shows the mass and spin values that the spacetime described by equation (\ref{metric}) needs to have
	in order to have a light-ring frequency of 250 Hz.
This frequency is broadly consistent with the ringdown frequency of GW150914 (\cite{TheLIGOScientific:2016src}).
It can be seen from the figure that for values of $b$ greater than zero,
	either a higher mass or a lower frequency is needed obtain the same frequency as the required values in general relativity.

For values of $b$ greater than $b_{\mathrm{maxLR}}$ there is no light ring.
In this case the ringdown of the object will be dominated by its intrinsic quasi-normal modes and will look radically different from the damped-sinusoid ringdown associated with the light ring.
The exact calculation of this behaviour will likely require a numerical solution of the field equations of pc-GR and is beyond the scope of this paper.

\section{Pre-merger: inspiral}

In order to solve the two-body problem for the orbital motion of two nearly equal mass objects,
	more is needed than just the one-body metric (\ref{metric}).
In Newtonian gravity the two-body problem of bound gravitational orbits is solved by the Keplerian orbits.
These Keplerian orbits, along with the Einstein quadrupole formula for gravitational wave emission,
can be used to infer basic properties of the binary source of GW150914 (\cite{Abbott:2016bqf}).
Beyond this Newtonian order,
post-Newtonian (PN) corrections to the orbits can be calculated,
which impact the gravitational wave signal.
These are usually regulated (\cite{Cutler:1994ys, Blanchet:1995ez, Blanchet:2013haa})
	by the PN parameter $x \sim \left( v/c \right)^2$,
	the dimensionless spins,
	and the mass ratio $q = M_1/M_2$,
	with $M = M_1+M_2$ the total mass
	and $\mu = M_1 M_2 / M = M q / (1+q)^2$ the reduced mass.
To these we now add the parameter $b$ which regulates the relative strength of the modification to the function $\psi$.
In the Newtonian and post-Newtonian regimes, $x \sim M / r \sim (M\w)^{2/3}$.
We thus see from the factor $g(r)$ in (\ref{modified mass}) that every appearance of $b$ involves a suppression by at least $n$th pN order.
We shall approximate how the leading order correction to the Newtonian frequency and phase evolutions depends on the modification,
and find the $b$ dependent post-Newtonian term to leading order in $b$.

In the wave zone we expect the same relation between the metric perturbation and the source quadrupole as in GR,
	and so expect the same wave polarizations and multipole decomposition.
We therefore treat the generation of gravitational waves
	as governed by a quadrupole formula
	(\cite{Einstein:1918btx,Blanchet:1995ez,Blanchet:2013haa,Abbott:2016bqf}),
\beq
  \label{quadrupole}
  \dot{E}_{\rm GW}
  = - \frac{32}{5} \frac{G}{c^5} \mu^2 \, r^4 \, \w^6 \, g^\varrho(r)~,
\eeq
where we allow for a possible deviation from the GR quadrupole formula
	with a subleading term $g^\varrho(r)$.
Examining the energy carried by the waves far away suggests adopting $\varrho=0$,
	while we note that \cite{Hess:2016gmh} uses $\varrho=1$ for regulating this emission.

The output of gravitational waves drains the orbital energy of the system,
    which is assumed to descend through quasi-circular orbits.
To leading order in $b$ this orbital energy is
\ba
\label{Eorbital}
E_{\rm orb} &=& -\frac{G m_1(r) m_2(r) } {2r}				\nonumber \\
	&=& -\frac{G M \mu } {2r} \left[ 1
										- b \left( \! \frac{M}{r} \! \right)^{\!n} \!\! Q
										+ b^2 \left( \! \frac{M}{r} \! \right)^{\!2n} \!\! Q_2
								\right],~~~~~~
\ea
where $Q = \frac{1 \!+\! q^n} {\left( 1 \!+\!q \right)^n}$,
and $Q_2 = \frac{q^n} {\left( 1 \!+\!q \right)^{2n}} \mathfrak{f}(n,\eps_1, \eps_2)$
depend on the mass ratios and distributions\footnote{
The form-factor
\ba
\mathfrak{f}(n, \eps_1, \eps_2) = n 	\sum_{k=1}^{n\!+\!2} (-1)^k
		\frac{\Gm(n\!+\!k\!-\!2)\,\Gm(n\!+\!2\!-\!k)}{\Gm(n\!+\!2)\,\Gm(n\!-\!1)}	\cdot		~~~~~~~~	\nonumber\\
\cdot
		\left. \left[ x^{-(n\!+\!2\!-\!k)}	
					\sum_{m=0}^1
							(-1)^{m(n-1)} (x\!+\!(-1)^m)^{-(n\!+\!k\!-\!2)}
		\right] \right|^{1-\eps_2}_{\eps1}
\!\!\!\!\!
\ea
also generally depends on the how the singularities near $r=0$ are regularized.
In the simplest model, $\mathfrak{f}(n, \eps_1, \eps_2) = 1$.
}.	% of \footnote
We note that $Q=1,\,Q_2=0$ corresponds to the model of \cite{Hess:2016gmh}.
We also introduce for the deviation from GR the shorthand
\beq
\gt = 1 - g(R) = b \left( \! \frac{M}{r} \! \right)^{\!n} ~~,~~
\frac{\dd \gt}{dr} = -\frac{n}{r}\gt ~,
\eeq
such that to leading order in the deviation,
\beq
\label{Kepler}
 \frac{\dd E_{\rm orb}}{\dr}
 = \frac{G M \mu }{2r^2}
	\left[	1 - (n\!+\!1) \, Q \, \gt	\, \right] \, ,
\eeq
which can be used in the equation for the energy balance equation
	$\dot{E}_{\rm GW} = \dot{E}_{\rm orb} = {E}_{\rm orb}'\dot{r}$
	to find
\beq
\label{energy balance short}
 - \frac{32}{5} \frac{G}{c^5} \mu^2 \, r^4 \, \w^6 \, g^\varrho(r) = \frac{G M \mu }{2r^2}
	\left[	1 - (n\!+\!1) \, Q \, \gt	\, \right] \dot{r}	~,
\eeq
We note that for large $b$, the gravitational well might have a minimum
	at finite $r = M ^{\,n} \!\!\!\! \sqrt{(n\!+\!1)\,b\,Q}$,
	by which the energy balance approximations fail.
Before reaching there, 
	the orbital angular velocity $\w$ can be eliminated
	from the expression (\ref{energy balance short}) by noting that,
    for quasi-circular orbits,
    there is a relation between $\w$ and $r$ given similarly to Kepler's third law by
\beq
\label{Kepler2}
\w^2 = \frac{G M }{r^3} \left[	1 - (n\!+\!1) \, Q \, \gt \,		\right] \, ,
\eeq
Thus we find from equation (\ref{energy balance short}) that
\beq
 \label{energy balance full r}
 \dot{r} = - \frac{64}{5} \frac{G^3 M^2 \mu}{r^3 c^5} \, 
	\left[	1 - (n\!+\!1) \, Q \, \gt \,	\right]^2 \, g^\varrho(r) ~,
\eeq
an equation which can be solved numerically for the orbit.
%which can be immediately integrated to
%\beq
%  \label{integral}
%r^4 \left[ 1 + 8 (n\!+\!1) \, \IBr \, \right]
% = \frac{256}{5} \frac{G^3 M^2 \mu}{c^5} ( t_c - t ) \,.
%\eeq
%For $n \in \{1,2,3,4,6,8,12,16\} $ we may solve analytically for $r(t)$, and then for $\w(t)$.

\subsection{Amplitude evolution}
\begin{figure}
	\includegraphics[width=\columnwidth]{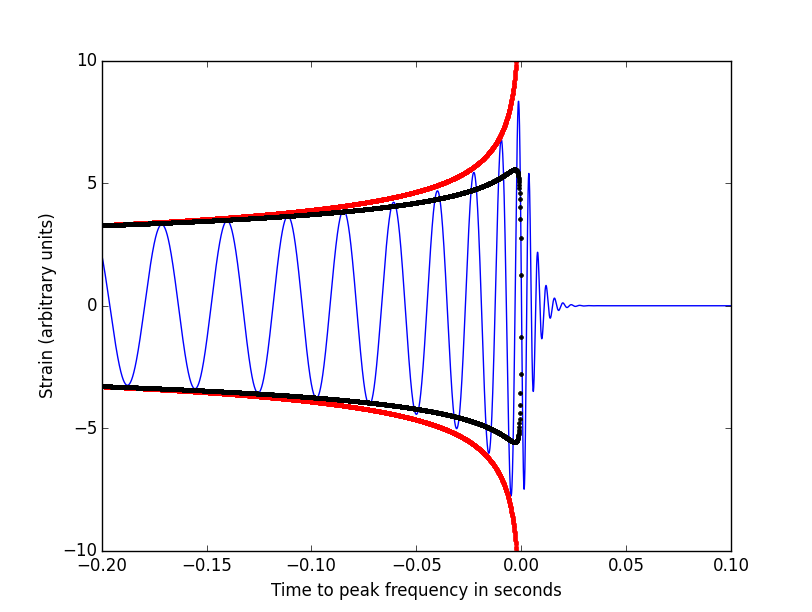}
    \caption{The amplitude evolution of the inspiral with $b=0$ to leading order 0PN approximation (red)
	and with $b=16/27$ (black) to leading order in the modification, for the $n=2$ case.
	These are compared to the general relativity model SEOBNRv2 full gravitational waveform (blue)
    for a binary coalescence of near equal mass objects each with zero spin.
    The waveform is shown from an initial frequency of 30Hz,
    similar to the low frequency cutoff of the Advanced LIGO detectors during their first observation run
    and the amplitudes are matched there.
    The $b=0$ 0PN approximation with $b=0$ is seen to match better than the $b=16/27$ approximation,
    especially for the later inspiral cycles, just before the peak.}
    \label{fig:amplitude}
\end{figure}

The evolution of the amplitude of the gravitational wave from a binary inspiral can be found
	by substituting $r(t)$ from eqn (\ref{energy balance full r}) and $\w(t)$ from eqn (\ref{Kepler2})
	into the Newtonian order amplitude equation (compare \cite{Hess:2016gmh}'s  Eq. (14), with $\varrho=1$)
\beq
A = \frac{4G\mu \w^2 r^2}{d_{L}c^4}g^\varrho(r)	~.
\label{amplitude}
\eeq
Since this relies on several approximations it is not expected to exactly match the amplitude evolution for a real signal.
However, it is known that the 0PN Newtonian amplitude is a good approximation to the full signal in general relativity
	up until very close to the merger (\cite{Cutler:1992tc}).
A plot of the amplitude evolutions in pcGR and the 0PN Newtonian approximation are shown in Fig. (\ref{fig:amplitude})
	against a full inspiral-merger-ringdown waveform model in vacuum general relativity, SEOBNRv2 (\cite{Taracchini:2013rva}).

It can be seen in the figure that the approximation based on $b=16/27$ begins to deviate noticeably from the GR model waveform several orbits before the peak amplitude.
A comparison of the measured gravitational wave amplitude against the predicted amplitude can constrain the luminosity distance to the source via equation (\ref{amplitude}).
However, the model should match the amplitude evolution for all values of the inspiral expansion parameter $M/r$ as the orbit evolves and the bodies approach one another.
It is not possible to compare amplitudes only at a single value of the expansion parameter as was suggested in \cite{Hess:2016gmh}.
Because pc-GR agrees with GR for sufficiently low values of $M/r$ where several cycles of the waveform are seen, the luminosity distance is required to be broadly consistent with the luminosity distance found by the LIGO team (\cite{Abbott:2016blz}),
	and the high redshift values found in \cite{Hess:2016gmh} for GW150914 are not consistent with the data,
	even within pseudo-complex general relativity.

\subsection{Phase evolution}

To obtain the phase evolution of the orbital motion and of the gravitational wave,
	it is useful to work in the frequency domain in the PN framework (to leading and next-to-leading orders).
Differentiating the Keplerian relation (\ref{Kepler2}) with respect to time, yields after some algebra
\beq
\label{w-r}
\left[ 3 - n (n\!+\!1) \, Q \, \gt \, \right] \frac{\dot{r}}{r} = -2 \frac{\dot{\w}}{\w} \, .
\eeq
and we can change from $r$ to $\w$ using eqns (\ref{Kepler2},\ref{w-r}),
\ba
r &=& \left[\frac{G M }{\w^2} \right]^{1/3}
		\left[	1 - \frac{(n\!+\!1)}{3} \, Q \, \gtw	\,	\right] \, ,																\\
\dot{r} &=& -\frac{2}{3}\frac{\dot{\w}}{\w} \, r \, \left[ 1 + \frac{n (n\!+\!1)}{3} \, Q \, \gtw	\, \right] \, ,
\ea
with
\beq
\gtw = \tilde{g} \left( r(\w) \right) = b \left( M \w\right)^{2n/3\,}	 ,	~~
g^\varrho(r) = 1- \varrho\,\gtw ~ .
\eeq
Hence instead of (\ref{energy balance full r}) we have
\ba
 \label{energy balance full w}
 \dot{\w} = \frac{96}{5} \frac{\left(G\Mc\right)^{5/3}}{c^5} \, \w^{11/3}
				\left[	1  - \IBw		\right]	 ,	\\
 \label{energy balance full f}
 \dot{f} = \frac{96}{5} \frac{ \pi^{8/3} \left(G\Mc\right)^{5/3}}{c^5} \, f^{11/3}
				\left[	1  - \IBw			\right]	 ,
\ea
as the new chirp equations, with the standard chirp mass $\Mc = \left( M^2 \mu^3 \right)^{1/5}$
and with $\gtw$ and numerical prefactors collected into the modification at $n$-PN $\IBw$,
\beq
\IBw = \left( \frac{(n\!+\!2) (n\!+\!1)}{3} Q + \varrho\right) b \,  (M\w)^{2n/3}~.
\eeq
We note also (\ref{energy balance full f}) in terms of the gravitational wave frequency
$f = \w/\pi$ (twice the orbital frequency),
with $\IBf = \IB (\w=\pi f)$.

We would like to relate this $\IBf$ modification to known bounds on PN coefficients.
We first plug equation (\ref{energy balance full f}) into the integrals for the time and for the phase (compare \cite{Cutler:1994ys})\footnote{
These forms must be trivially modified to apply to $n=2.5,4$
	where the integrals for $\phi$ and $t$ respectively give logarithms of $\pi M f$ rather than its powers.
},		% of \footnote{
\ba
t 	&=& \!\! t_c + \int \frac{\df}{\dot{f}}																					\nonumber\\
	&=& \!\! t_c - \frac{5 \, c^5 \, \left(\pi f \right)^{-8/3}} {256 \left(G\Mc \right)^{5/3}}
			\left[ 1 - \frac{4}{n-4} \IBf \right]	
 \, ,~~~~~~																												\\
\phi &=& \!\! 2\pi \int \! f \, dt = 2\pi \int \!\! \frac{f}{\dot{f}}  \df													\nonumber\\
	&=& \!\! - \frac{c^5}{16 \left(\pi G\Mc f\right)^{5/3}} \left[ 1 - \frac{5}{2n -5} \IBf \right]	
 \, ,~~~~~~
\ea
and then use these with the stationary phase approximation (\cite{Cutler:1994ys}) to find
\ba
\Psi
&=& \!\!\! 2\pi f t_c - \phi_c - \pi/4		\\
	&&\!\!\!+ \frac{3 }  {128 \left(\pi G \Mc f \right)^{5/3} }
		\left[ 1 + \frac{20}{\left(n -4\right)\left(2n -5\right)} \IBf \right]	\, .	\nonumber
\ea

This form can be compared directly to the expected PN coefficients in GR of \cite{Buonanno:2009zt}
	(following \cite{Iyer:1993xi, Will:1996zj, Blanchet:2000nv}),
	and to the limits set on deviations from them
	by the observed gravitational waves in the inspiral regime in \cite{TheLIGOScientific:2016pea, TheLIGOScientific:2016src}
	(based on \cite{Talmadge:1988qz, Mishra:2010tp, PhysRevD.85.082003}).
This comparison is summarized in Table \ref{PN table},
	for the leading pc-GR PN terms for $n=1,2,3$ and the corresponding GR PN phase coefficients of orders $1,2,3$.
All coefficients are calculated for the fiducial equal mass non-spinning case ($q=1$, $a=0$);
	the pc-GR coefficients are calculated for the critical $b_{\mathrm{maxH}}$ value of equation (\ref{bmaxH}), where the horizon vanishes,
	and for $\varrho=0$.
The table also compares to the 90\% bounds set on the relative deviations $(p_n^{\rm mod-GR} - p_n^{GR})/p_n^{GR}$
	established in LIGO's first observation run O1.
This comparison provides independent evidence to show that non-existence of horizons for $n=1$ pc-GR
	is inconsistent with observed gravitational wave events in O1,
	and the first evidence to rule out their possibility in $n=2$ pc-GR as well.

%%%%%%%%%%%%%%%%%%%%%%%%%%%%%%%%
\begin{table}
  \centering \caption{pc-GR PN coefficients}
\begin{center}
\begin{tabular}{cccccccc}
  \hline
  $n$	& $\rmin$	& $b_{\rm crit}$	& $p_n^{\rm pc-GR}$	& $p_n^{\rm GR}$	& $\delta_\phi$	& $range(\delta_\phi)$ \\  \hline
  1		& 1			& 0.5				& 20/9						& 6.44					& $34\%$			& $(-20\%, 5\%)$	\\
  2		& $4/3$		& 16/27			& 320/27					& 46.2					& $26\%$			& $(-130\%, 15\%)$ \\
  3		& 1.5			& 27/32 			& -225/8					& -652					& $4.3\%$		& $(-100\%, 600\%)$	\\
\hline
\label{PN table}
\end{tabular}
\end{center}
\end{table}
%%%%%%%%%%%%%%%%%%%%%%%%%%%%%%%%

Conversely, the limits on the deviations of PN coefficients can be translated to limits on $b$,
	which for $n=1,2,3$ are $|b|\leq 0.85, 2.96, 118$ respectively.
This suggests that $b \left(\pi M f \right)^{2n/3}$ is indeed a small parameter throughout the system's evolution,
	and that hence the introduction of the pc-GR modifications should not produce large deviations from the
	standard GR post-Newtonian inspiral, and in particular should not largely affect the chirp mass $\Mc$.
For the GW150914 data, estimating the chirp mass from directly from $f$ and $\dot{f}$ at different inspiral times
	using the Newtonian approximation (0PN, $b=0$)
	shows it remains approximately constant up to a frequency of $\sim150Hz$ (\cite{Abbott:2016bqf}),
	and is equal to roughly 30 solar masses.
This corresponds to $\left(\pi M f \right)^{2/3} \sim 0.17$,
	consistent with treating $\IBf$ only at leading order,
	and inconsistent with the much larger modified chirp mass (and correspondingly higher redshift)
	estimated in \cite{Hess:2016gmh} from the late part of the orbit.

\subsection{Testing a previous model}
As mentioned through the text, the model of \cite{Hess:2016gmh} can be considered under our formalism
	as the case $n=2$, $Q=1$, $Q_2=0$, $\varrho=1$.
For the critical value of $b=16/27$,
	this changes the leading coefficient $p_2^{\rm GR}$ from $320/27$ to $800/27$,
	which changes $\delta_\phi$ from $26\%$ to $65\%$ of the GR value.
As Table \ref{PN table} indicates,
	this value is well beyond the range observed and reported by LIGO's O1,
	and so this model pc-GR also cannot sustain horizonless objects as the sources of LIGO's detections.

\section{Conclusions}

We have shown how the model of pseudo-complex general relativity can be constrained
	using gravitational wave observations.
These observations are very much independent of and complementary to observations
	with other techniques that have previously been proposed, 
	such as accretion disk studies and imaging of super-massive black holes.
For two merging compact objects,
	gravitational wave observations provide strong constraints on the near horizon behaviour
	from both the inspiral phase and the final ringdown after merger.
In particular we have seen that the ringdown phase,
	modeled in terms of the light ring structure within pc-GR,
	typically requires the final object after merger to be slightly heavier and spinning slightly slower
	than is found using Einstein's GR.
However, the Newtonian limit of pc-GR requires the chirp mass
	to be broadly consistent with the values found using Einstein's GR,
	and this bounds both the total mass and luminosity distance
	to be broadly consistent with those found using Einstein general relativity.

We have discussed the model in terms of a dimensionless parameter $b$ and power index $n$ in equation (\ref{modified mass}).
For sufficiently large values of $b$ objects can be horizonless.
We find that horizonless objects in the $n=1$ case are already ruled out,
	independently of other Solar System constraints (\cite{Will2006}).

For the critical case with $n=2$ and $b=16/27$ discussed in \cite{Hess:2016gmh}
	we find that the final object can only fit the ringdown with a frequency in the range observed in \cite{TheLIGOScientific:2016src} if the spin value is close to zero.
Larger spin values require lower values of $b$ and hence, if $b$ is universal,
	allow non-spinning black holes with horizons.
In the inspiral regime, the case of $n=2$ and $b=16/27$ is also in tension with the data.
The model suggests a noticeable fall-off in the amplitude which is not seen in the data \cite{Abbott:2016bqf},
	and the inspiral phasing is in conflict with detailed fits of the post-Newtonian parameters (\cite{TheLIGOScientific:2016pea}).
	
For values of $n$ larger than 2 the situation is not so clear.
As the value of $n$ is increased in the model, the corrections of the model are constrained to smaller and smaller distances.
For $n=3$ the leading order correction from pseudo-complex general relativity is a 3PN term and this term is less tightly constrained by current LIGO observations.
Higher terms at 4PN and beyond are not yet fully calculated in general relativity so a direct comparison with these terms is not yet possible.  

Our conclusions are only valid to the extent of the approximations that have been made in deriving the model of \cite{Caspar:2012ux}.
The parameters $b$ and $n$ are assumed constant and apply equally to the pre-merger inspiral phase as to the post-merger ringdown.
Little is known about how pseudo-complex general relativity behaves in the highly dynamical merger phase and our results cannot address this part of the LIGO observations.
This is ultimately likely to require, as in the case of vacuum Einstein relativity, numerical solutions to the field equations.
Further work is also required to understand exactly how the features of pseudo-complex coordinates should be implemented in a gravitational theory, but this is beyond the scope of this work.

We have deliberately interpreted the pseudo-complex relativity model slightly differently from \cite{Hess:2016gmh}.
We have explicitly mapped the vacuum solution of pseudo-complex general relativity to an equivalent problem of what would be a non-vacuum spacetime in Einstein relativity.
This enables us to relate the techniques more generally.
Aside from pseudo-complex general relativity,
	this analysis can also be applied to dirty black holes,
	and a similar analysis of PN effects has been pursued for dark matter minispikes (\cite{Eda:2014kra}).
Thus it is hoped that the techniques here described may find application beyond the specific scope of testing pseudo-complex general relativity.

\section*{Acknowledgements}

We thank P. O. Hess and the LIGO TestingGR working group for useful conversations and correspondance.

%%%%%%%%%%%%%%%%%%%%%%%%%%%%%%%%%%%%%%%%%%%%%%%%%%


%%%%%%%%%%%%%%%%%%%%%%%%%%%%%%%%%%%%%%%%%%%%%%%%%%

\bibliographystyle{mnras}
\bibliography{references}

% Don't change these lines
\bsp	% typesetting comment
\label{lastpage}
\end{document}